\documentclass[12pt,preprint]{aastex}
\usepackage{graphicx}% Include figure files
\usepackage{bm}      % bold math
\usepackage{color}
\newcommand{\fei}{{\tt FeI} }
\newcommand{\runea}{{Run E} }

\newcommand{\runca}{{Run C} }
\newcommand{\runngrey}{{Run C-NG}}
\newcommand{\runngreya}{{Run C-NG} }
\newcommand{\runprandtla}{{Run G-P} }

\newcommand{\runultraa}{{Run G} }

\newcommand{\runhypera}{{Run H} }
\newcommand{\hinodea}{{\sl Hinode} }
\newcommand{\hinode}{{\sl Hinode}}
\newcommand{\stoproa}{\texttt{STOPRO} }
\newcommand{\murama}{\texttt{MURaM} }

\newcommand{\pdfa}{{PDF} }
\newcommand{\pdf}{{PDF}}
\newcommand{\pdfsa}{{PDFs} }
\newcommand{\pdfs}{{PDFs}}
\newcommand{\blappa}{{$B^{\mbox{L}}_{\mbox{app}}$} }
\newcommand{\blapp}{{$B^{\mbox{L}}_{\mbox{app}}$}}

\newcommand{\Phiu}{{\left< |B_z| \right>}}
\newcommand{\Phiua}{{\left< |B_z| \right>} }

\def\AD#1{{\textcolor{blue}{\bf #1}}} % TEXT REVISED AFTER REVIEW
\def\AD#1{#1} % EDITORS: uncomment this line!!!
\def\resp#1{{\textcolor{red}{\bf #1}}} % TEXT REVISED AFTER ACCEPTANCE
\def\resp#1{#1} % EDITORS: uncomment this line!!!
\slugcomment{}%30 Nov 2008}%submitted to ApJ.}

\shorttitle{Turbulent magnetic fields in the quiet Sun}
\shortauthors{Pietarila Graham, Danilovic, and Sch\"ussler}

\begin{document}


\title{Turbulent magnetic fields in the quiet Sun: implications of {\sl Hinode} observations and small-scale dynamo simulations}

%% from  v4.0. You can use \email to mark an email address

\author{Jonathan {Pietarila Graham}, Sanja Danilovic, and Manfred Sch\"ussler}
\affil{Max-Planck-Institut f\"ur 
Sonnensystemforschung, 37191 Katlenburg-Lindau, Germany}
%\email{jpietarilagraham@mailaps.org}

\begin{abstract}
Using turbulent MHD simulations (magnetic Reynolds
numbers up to $\approx8000$) and {{\sl Hinode}} observations, we
study effects of turbulence on measuring the solar magnetic
field outside active regions.  Firstly, from synthetic Stokes~$V$ profiles for the {{\tt FeI}
}lines at 6301 and 6302 \AA, we show that a peaked probability
distribution function (PDF) for observationally-derived field
estimates is consistent with a monotonic PDF for actual vertical
field strengths.  Hence, the prevalence of weak fields is greater than
would be naively inferred from observations.  Secondly, we employ the
fractal self-similar geometry of the turbulent solar magnetic field to
derive two estimates (numerical and observational) of the true mean
vertical unsigned flux density.  We also find observational evidence
that the scales of magnetic structuring in the photosphere extend at least down to \AD{an order of magnitude smaller than
$200\,$km:} the self-similar power-law scaling in the signed
measure from a {{\sl Hinode}} magnetogram ranges (over two decades in
length scales and including the granulation scale) down to the $\approx200\,$km resolution limit.  From the
self-similar scaling, we determine a lower bound for the true quiet-Sun mean
vertical unsigned flux density of $\sim50\,$G.  This is consistent
with our numerically-based estimates that 80\% or more of the vertical
unsigned flux should be invisible to Stokes$-V$ observations at a resolution of $200\,$km owing to the cancellation of signal from opposite magnetic polarities.  Our
estimates significantly reduce the order-of-magnitude discrepancy
between Zeeman- and Hanle-based estimates.
\end{abstract}

%% example has been keyed in  style. See the instructions to authors

\keywords{Sun: magnetic fields --- turbulence --- MHD --- techniques: polarimetric}

%% In the first two sections, notice the use of the  \citep
%% tied to the reference list via symbolic . The KEY corresponds

%% Note that for sources with brackets in their names, e.g. [2004] 14h-090,

\section{Introduction}

Determining the strength of the magnetization of the ``quiet''
  Sun is tied to the question of how much flux resides at small
  scales.  This is important, for example, in determining the energy
  budget available for chromospheric heating.  Observationally, two
  methods are employed to constrain the solar magnetic field: the
  Hanle and Zeeman effects.  The Hanle effect measures (in principle)
  the mean magnetic field strength, $\left<|B|\right>$, as there are
  no cancellation effects, but quantitative interpretation requires
  assumptions about the probability distribution function (\pdf) of
  the turbulent magnetic field.  The Hanle de-polarization is
  measured in strong\resp{er} lines formed in the \resp{mid- to upper-}photosphere and
  estimates are made of $\left<|B|\right> \sim 130\,$G, the field residing
  primarily in the intergranular lanes \citep{TrBuShAsRa2004}.  The
  Zeeman effect measures the longitudinal component (via Stokes~$V$) and
  transverse components (via Stokes~$Q$ and $U$) of the magnetic field but
  suffers from cancellation effects.  Hence, it is ``blind'' to any
  ``hidden'' mixed-polarity flux at scales smaller than the resolution limit of an
  instrument.  However, the benefit of Zeeman measurements is that their interpretation
  requires no assumption about the turbulent \pdfa and measurements
  can actually be used to determine the \pdfa on scales larger than the
  resolution limit.  Some attempts to incorporate the effects of
  cancellation into Zeeman-based Stokes inversions utilize the
  micro-structured field hypothesis \citep{SaAlLadeMaPi+1996}.
  However, typical estimates based on the longitudinal Zeeman effect give
  $\left<|B_z|\right>\sim10\,$G for the
  mean unsigned vertical flux density (see Table 3 in
  \citealt{BeGoYeChOk+2008} for a review of the spread of recent
  Zeeman-based estimates).  These values are significantly smaller
  than the Hanle-based estimates.
  %However, synthetic Stokes $V$ observations
  %from magnetohydrodynamic (MHD) simulations of photospheric
  %magnetoconvection with a mean magnetic field strength
  %$\left<|B|\right> \approx 20\,$G, a resolution of $20\,$km$\,\approx
  %0\arcsec.03$, and no dynamo action compare well with Zeeman-based
  %observations \citep{KhShSo+2005,KhMaGoCo+2005,BeGoYeChOk+2008}.
  The discrepancy between the results from Hanle ($\sim100\,$G) and
  Zeeman ($\sim10\,$G) measurements is not unexpected since the Zeeman
  observations see only the resolved flux while the Hanle interpretation
  depends on assumptions about an unknown \pdf.

In this work, we ask the question ``How can the turbulent
  fractal geometry of the magnetic field in the solar photosphere be
  accounted for in interpretations of Zeeman-based observations?''  We
  attempt to derive the spectral/fractal properties from high
  resolution ($0\arcsec.3$) Zeeman observations with the
  \hinodea spectro-polarimeter (SP)
  \citep{KoMaSa+2007,TsIcKa+2008,LiElSt2001} and high resolution (down
  to $4\,$km$\,\approx 0\arcsec.006$), turbulent dynamo
  simulations (up to magnetic Reynolds numbers, $Re_M \approx 8000$)
  with the \murama code \citep{V03,VSS+05}.  First, we clarify the
  consistency of \pdfsa derived from observations and simulations,
  respectively.  Zeeman observations typically show \pdfsa of
  quantities derived from Stokes~$V$ which can be described as a peaked
  function \citep{KhMaGoCo+2005,LiKuSoNa+2008}.  We demonstrate that
  the peaked \pdfsa from Stokes~$V$ measurements and the monotonic \pdfsa
  typically reported from numerical simulations are, in fact,
  compatible.  Failure to take into account this observational
  bias can lead to a gross underestimation of the prevalence of
  weak fields and, consequently, to incorrect estimates of the mean
  magnetic field strength and magnetic energy density in the solar
  photosphere.

Second, we extrapolate the results obtained with observations and
simulations (based on their respective fractal properties) to scales
below the resolution limits to estimate the amount of ``hidden'' flux.
To do this, we exploit the fact that the fractal geometry is intimately
connected to power-law scaling relations such as
  $N(l)\propto~l^{-D_f}$.  Here, $N(l)$ is the number of boxes of edge
  length $l$ covering a fractal set (such as all pixels with apparent
  magnetic flux above some threshold) and $D_f$ is the fractal
  dimension.  It has long been known that the distribution of plage
magnetic field is such a statistically self-similar fractal
\citep{ScZwBa+1992,BaScZw+1993}.  As the threshold can influence the
fractal dimension inferred, however, the geometrical structure is more
complicated than a simple fractal.  In this case, the fractal concept
must be generalized by adding a measure defined by the absolute
  value of the net magnetic flux through each box of edge length $l$.
This measure also displays self-similarity (a power-law scaling)
for the solar quiet Sun network magnetic flux (down to
$0\arcsec.5$ resolution in \citealt{LaRuCa1993} and
\citealt{CaLaRu+1994}; see also \citealt{KrSo2004}) and also
for numerical simulations of magnetoconvection \citep{BrPrSe+1992}.
Hence, the geometry of the magnetic field is said to be multifractal.

Cancellation effects do not play an important role in observations of the
multifractal unipolar solar regions discussed above.  In contrast, if
the magnetic field of the quiet Sun internetwork is multifractal (as
we will show), cancellation can play a significant role for scales
below $1\arcsec$ \AD{(see also \citealt{SaAlEmCa2003}).}  For these mixed-polarity fields it is necessary to
  further generalize the fractal concept to a signed measure.  Here,
  the power-law scaling exponent is the cancellation exponent
  \citep{ODS+92}.  If this self-similar power-law scaling extends below the
  observational resolution, small scale cancellation will occur and
  correct values of the mean field strength and energy cannot be
  established.  As was pointed out by \citet{LaRuCa1993}, such a signed measure could
  also be employed to extrapolate moments of the magnetic field
  below resolvable limits. Until now, no attempt has
  been made to employ self-similar scaling to estimate the total
  cancellation, and, hence, the true mean quiet-Sun magnetic field
  strength.
  
%Our analysis demonstrates that cancellation does continue
%self-similarly to scales smaller than $200\,$km and additionally
%allows for estimates of the unsigned vertical flux (flux density times
%area of the observed region) that remains undetectable at this
%resolution.

\section{\AD{Data and Methods}}

We use \murama simulations \citep{V03,VBS04,VSS+05} for a rectangular
domain of horizontal extent $4.86 \times 4.86\,$Mm$^2$ and a depth of
$1.4\,$Mm.  Runs of small-scale, local dynamo action with increasing resolution and Reynolds numbers
have been carried out (see Table \ref{TABLE:RUNS}).  The turbulence is
sufficient for small-scale dynamo action
\citep{VoSc2007}.  All magnetic field in the simulations results
from dynamo amplification of a small seed field as there is no net flux through
the box, no flux advected into the box, and no initial large-scale field. A well-known rule-of-thumb
for forced turbulence simulations is that one decade of scales smaller
than the forcing is required before the inertial range begins and
another decade is required (at the opposite end of the spectrum) for the dissipative scales.  For
photospheric magneto-convection simulations, where the forcing granulation
scale is 1 Mm, this means that a grid resolution of 10 km or smaller
is required before a simulation becomes turbulent.
\AD{(See \citet{SaAlEmCa2003} for a study with a 10km resolution
using a Boussinesq simulation and multi-component Milne-Eddington line
synthesis.)}
  As we
compute simulations down to a grid resolution of 4 km, we are now able
to measure the effects of turbulence on observational quantities.

To this end, we require
synthetic profiles (for the \fei lines at 6301.5~\AA~and 6302.5~\AA) as
calculated from a snapshot of a run with non-grey radiative transfer.
Owing to the computational expense of such a run, it was started from a
snapshot from the statistically stationary state of \runca (see Table \ref{TABLE:RUNS}) and then run for approximately one convective turnover time, 10 minutes.
Stokes $V$, $Q$, and $U$ profiles were then computed in one dimension (1D)
assuming
local-thermodynamic-equilibrium (LTE) and using the \stoproa code in the
\texttt{SPINOR} package \citep{So1987,FrSoFl+2000}.  %A recent analysis has given
%an in-depth description of spectra obtained from \murama simulations
%in this way (\citealt{BeGoYeChOk+2008}; see also \citealt{ShScSo+2007}).
We will concentrate in this work, however, on Stokes~$V$ observations (and
  synthetic observations) at disk center.  This allows us to avoid
  line-of-sight effects, to identify the longitudinal component of the
  magnetic field as its vertical component, and to avoid the difficult
  estimation of the cancellation properties of Stokes~$Q$ and $U$.

Observational data is obtained from the spectro-polarimeter (SP,
\citealt{LiElSt2001}) of the Solar Optical Telescope (SOT,
\citealt{TsIcKa+2008}) on the \hinodea satellite \citep{KoMaSa+2007}.
\AD{One set, a spatial map,}
consists of $2048$ scans taken on March 10, 2007 (11:37:36 -- 14:36:48
UT), in the ``normal mode'' (exposure time of $4.8\,$s) with the
scanning step of $0.1476\arcsec$ and pixel size along the slit of
$0.1585\arcsec$. The data set covers a quiet Sun region at the
disk center, over the large field of view of $324\arcsec\times
164\arcsec$. \AD{The second set, a ``deep mode'' time series,
consists of a $103$ steps at disk center, each with an effective exposure time of $67.2\,$s after
application of a temporal running mean, and was taken on
February 27, 2007 (00:20 -- 02:20 UT).  Both data sets have previously
been described in \citet{LiKuSoNa+2008}.} Corrections for various instrumental effects are made
using the \texttt{SolarSoft}\footnote{http://www.lmsal.com/solarsoft/} procedure \texttt{sp\_prep} which calculates the wavelength-integrated Stokes~$V$ ($V_{\mbox{tot}}$),
\begin{equation}
V_{\mbox{tot}} = \mbox{sgn}(V_b) \frac{\big{|}\int_{\lambda_b}^{\lambda_0}V(\lambda)d\lambda\big{|}
+ \big{|}\int_{\lambda_0}^{\lambda_r}V(\lambda)d\lambda\big{|}}{I_c\int_{\lambda_b}^{\lambda_r}d\lambda}\,,
\label{eq:vtot}
\end{equation}
where $\mbox{sgn}(V_b)$ is the sign of the blue peak, $\lambda_0$ is the line
center, and $\lambda_{r,b} = \lambda_0 \pm 30\,$pm \AD{\citep{LiKuSoNa+2008}.}
The procedure includes
also a Milne-Eddington-based calibration of the sum of $V_{\mbox{tot}}$
from both lines into a measure of
longitudinal ``apparent flux density'',
$B^{L}_{app}$
\citep{LiIcKu+2008}. This
calibration is tailored to retrieve the field value from weak and
noisy internetwork signals. It assumes that magnetic structures
are spatially resolved (fill the resolution element) and does not
take into account the magnetic field variations over the height
range where lines are formed. Consequences of the latter assumption are
studied in the next section, when $B^{L}_{app}$, obtained from
Stokes~$V$ profiles synthesized from simulations, is compared with
the vertical component of the actual magnetic field.

\section{Results}

\subsection{\AD{\pdfs}}
\label{SEC:PDFS}

\begin{figure}
   \plotone{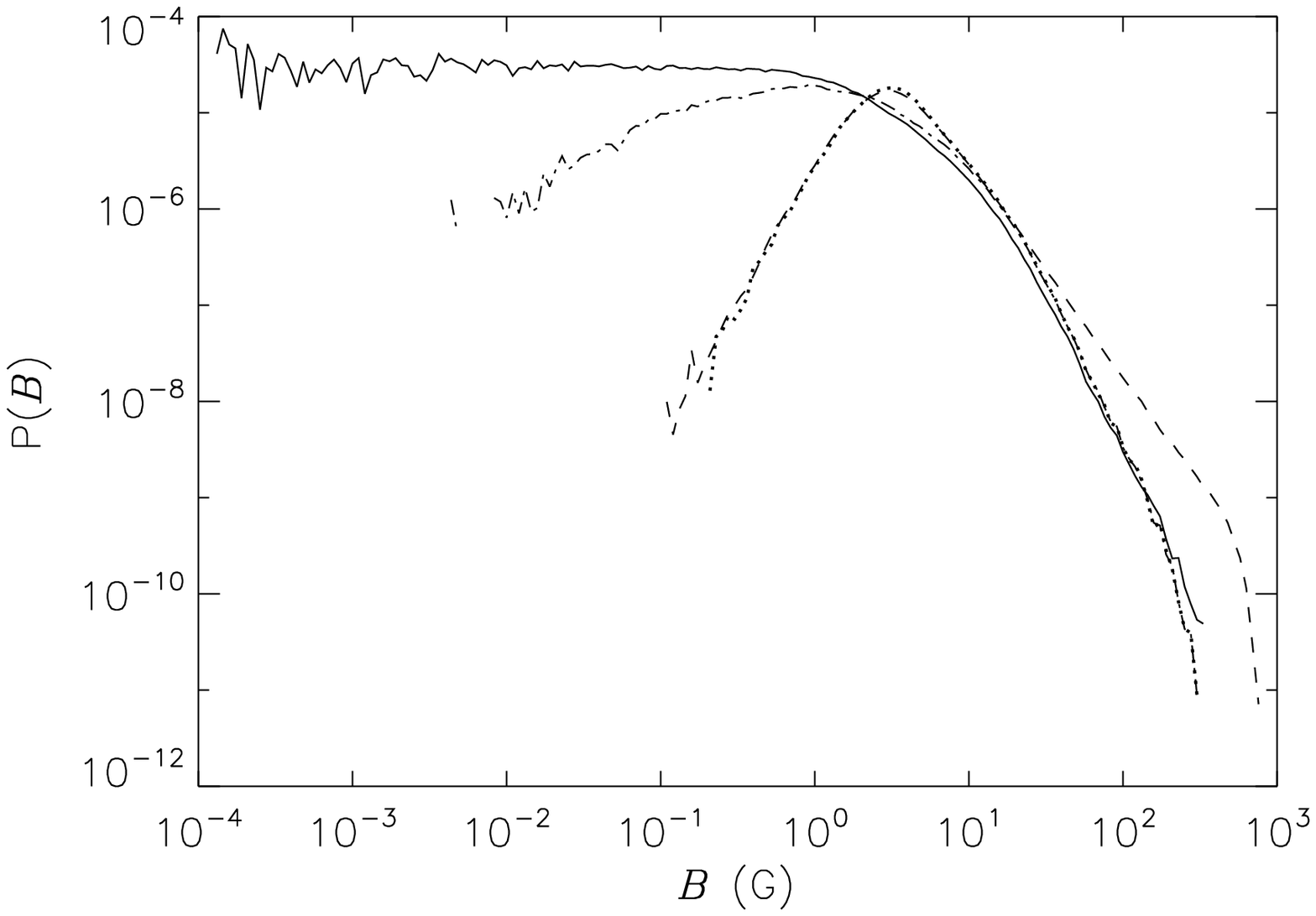} 
%  \plottwo{fig_pdfs_blapp}{fig_pdfs_bave}  
  \caption{Probability distribution functions (\pdfs) for magnetic
    field strengths and derived field proxies: \hinodea SP \AD{``normal mode'' map} \blappa
    (\AD{dashed} line), \murama simulation $B_{\mbox{ave}}$ (see text,
    solid line), 
    \murama synthetic \blappa (B derived from Stokes~$V$, \AD{dot-dashed}), and \blappa including \AD{a noise level of $1.1\times10^{-3}$ (dotted).}  The
    \pdfsa of the synthetic observations appear peaked although we
    have a monotonic distribution of vertical field strengths.
    %Failure to consider these effects can lead to a gross
    %underestimation of the distribution of weak fields.
%   We find the mean fields to be
%    $\left<|B^{\mbox{L}}_{\mbox{app}}|\right>=6.9 Mx cm^{-2}$ and
%    $\left<B_{\mbox{ave}}\right>=5.5\,$G.  This represents an over estimation
%    of 26\% due to the differences in the \pdfs.  {The average \pdfa for
%all \hinodea
%      $7\arcsec.2 \times 7\arcsec.2$ subregions with mean \blappa below $10\,$G
%      is also shown (dotted line).  Here the effect is larger than for
%the simulations--possibly due to stronger magnetic and velocity field gradients.}}
}
  \label{FIG:PDFCOMPARE}
\end{figure}

A marked difference exists between
the \pdfsa inferred from Zeeman polarimetry
(e.g., \citealt{KhMaGoCo+2005,LiKuSoNa+2008}) and the
\pdfsa from numerical computations (e.g., \citealt{SoNaSaAl2003,VoSc2007}).  For example, in Fig. \ref{FIG:PDFCOMPARE} we present
  the \pdfsa \AD{from the \hinodea ``normal mode'' map} magnetogram (apparent vertical magnetic flux density, \AD{dashed} line) and
  of \murama simulation \runngreya (average vertical magnetic field, solid line). The
  polarimetric observation peaks at $B^{\mbox{L}}_{\mbox{app}}\approx3\,$G
  while the simulation possesses a monotonic
  distribution without a distinct maximum: there exists a greater amount
of weak vertical field than indicated by the observations.  The observation also shows greater intermittency
  (the distribution has an enhanced strong signal (field) tail when
  compared to a Gaussian) than the simulation.  This can possibly be attributed to
  the much lower Reynolds number of the numerical simulation compared
  to the Sun
as well as to the smaller simulation box: contributions
from dynamo action in the deeper layers and supergranular network flux
concentrations are absent.
To address the qualitative difference
  (peaked versus monotonic) between simulations and
observations, we ask what \pdfa of
Zeeman-based observational signatures would result if \pdf$(B_z)$
monotonically decreases with increasing vertical field strength
(instead of possessing a peak).  Our approach will be to
assume the distribution of field strengths from turbulent \murama
simulations and examine the consequences
of such a distribution on Stokes~$V$ observations.

\begin{figure}
  \plotone{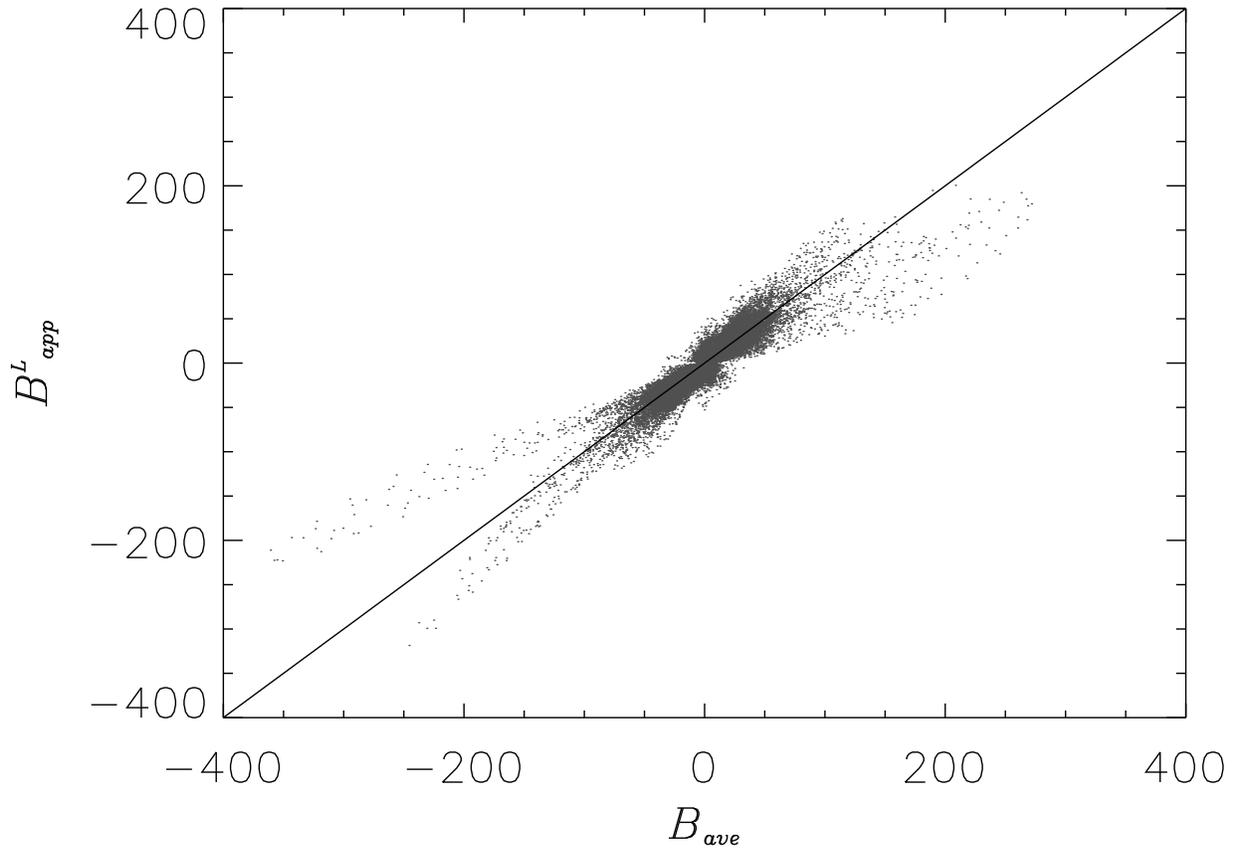}  
  \caption{\blappa derived from \murama \runngreya
    versus $B_{\mbox{ave}}$, the actual vertical magnetic field strength averaged over
    $\log\tau\in[-3.5,.1]$.  The linear Pearson correlation for the
    two quantities is $r=0.92$.  Note the large scatter.}
  \label{FIG:SNAPS}
\end{figure}

Though noise, resolution, and other instrumental factors are important
in any real observation, we first address the question assuming a
``perfect'' instrument.  Using the synthetic profiles from \runngrey,
we calculate $V_{\mbox{tot}}$ with Eq. (\ref{eq:vtot}) and determine \blappa following \citet{LiKuSoNa+2008}.  In Fig. \ref{FIG:SNAPS}, the derived \blappa signal versus
$B_{\mbox{ave}}$,
the vertical magnetic field strength averaged over the height range
corresponding to
$\log\tau\in[-3.5,.1]$,
%\begin{equation}
%B_{ave}(x,y) \equiv \int_{z\ni\log\tau \in [-3.5,.1]}B_z(x,y) dz\big{/}\int_{z\ni\log\tau \in [-3.5,.1]}dz
%\end{equation}
is shown.  This quantity was selected for its
linear Pearson correlation with \blappa of $r=0.92$ and its
coefficient of linearity, $B^{\mbox{L}}_{\mbox{app}} \approx 1.0 B_{\mbox{ave}}$, which
is consistent with the calibration of \citet{LiKuSoNa+2008}.  This height range also encompasses most of the formation height of the
  \texttt{FeI} lines at 6301 and 6302~\AA.  Though \blappa and
  $B_{\mbox{ave}}$ are well correlated, there is a large scatter.  We note,
  also, that changing the range to $\log\tau\in[-2,.1]$ does not
  significantly affect the correlation, $r$.  This indicates that most
  of the Stokes~$V$ signal is generated in deeper layers \citep{OrSuBeRudToIn2007}.

In Fig. \ref{FIG:PDFCOMPARE}, we present a comparison between the \pdfsa of
\blappa as derived from the synthetic Stokes~$V$
profiles (\AD{dash-dotted} line) and $B_{\mbox{ave}}$ (solid line).
\pdf$(B_{\mbox{ave}})$ monotonically decreases with increasing field strength
while \pdf(\blapp) shows a peak near $1\,$G and a strong decline
towards smaller field strengths.  The
\pdfsa for maximum Stokes~$V$ amplitude and total circular polarization
are qualitatively similar to that shown for \blapp.
\pdfsa for the vertical magnetic field from different volumes
and 2D cross sections from all simulation runs, chosen either by height or by optical depth,
show similar \pdfsa to that shown for $B_{\mbox{ave}}$.
That is, the monotonically decreasing distribution is a robust feature
of the vertical magnetic field when sampled by geometrical height, optical depth, or by
averaging over the vertical direction.
  The difference between observations and simulations is
  caused by the radiative transfer that produces circular polarization
  from longitudinal magnetic field. 

The above result shows that caution is needed when interpreting the
distribution of Stokes~$V$ signal in order to avoid a drastic
  underestimation of the occurrence of weak field.  This caution
naturally extends to moments of the distribution such as mean vertical
flux density or mean vertical magnetic energy density. For example,
\blappa and $B_{\mbox{ave}}$ are very well correlated with a coefficient of
linearity of unity, but their averages,
$\left<|B^{\mbox{L}}_{\mbox{app}}|\right>=6.9\,$G and
$\left<|B_{\mbox{ave}}|\right>=5.5\,$G, differ significantly.  We see that an
over estimation of 26\% results from assuming the vertical magnetic
field to have the same distribution as the signal derived from Stokes~$V$,
even in the absence of noise \AD{(note that this is close to the 20\%
loss found in \citealt{SaAlEmCa2003}).}

\begin{figure}
  \plottwo{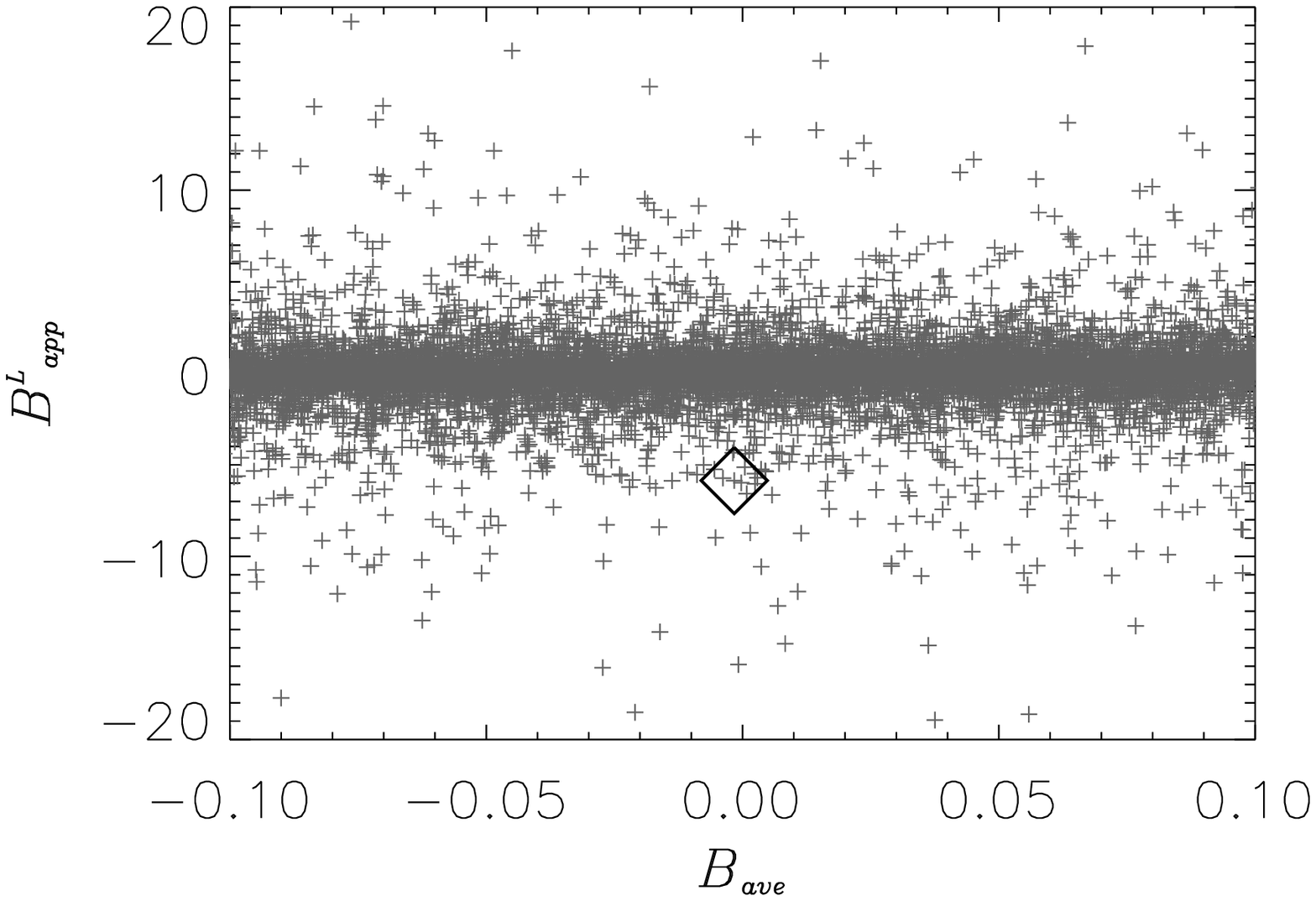}{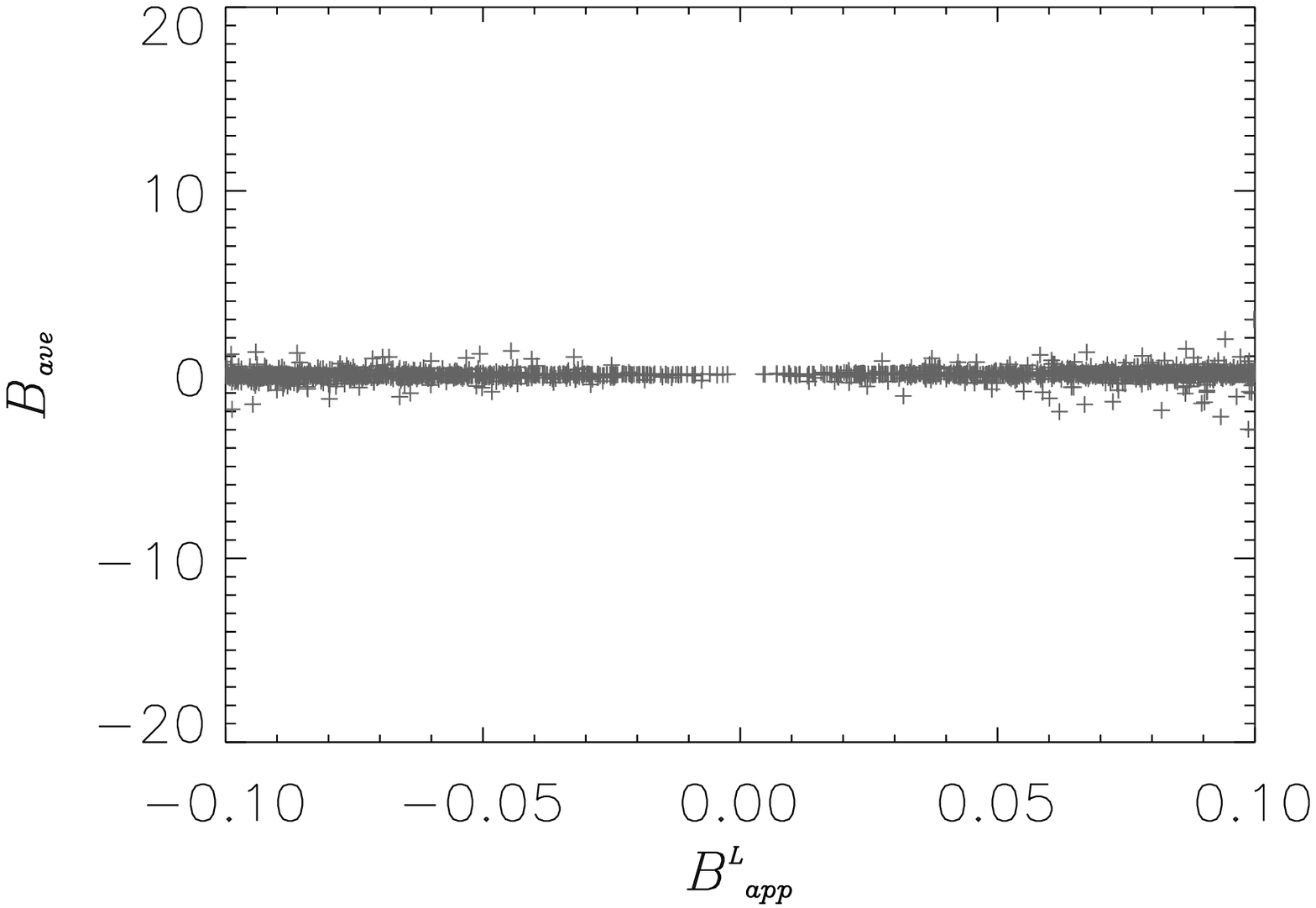}  
  \caption{{\bf (Left)} \blappa versus $B_{\mbox{ave}}$ for $B_{\mbox{ave}}<0.1\,$G {\bf
      (Right)} $B_{\mbox{ave}}$ versus \blappa for \blappa$<0.1\,$G (\blappa computed from noiseless V-profiles).  These plots
    indicate the bias that strong Stokes~$V$ signal can be associated
    with a pixel with weak averaged magnetic field, but seldomly {\sl
      vice-versa}.}
  \label{FIG:SNAPB}
\end{figure}

\begin{figure}
  \plottwo{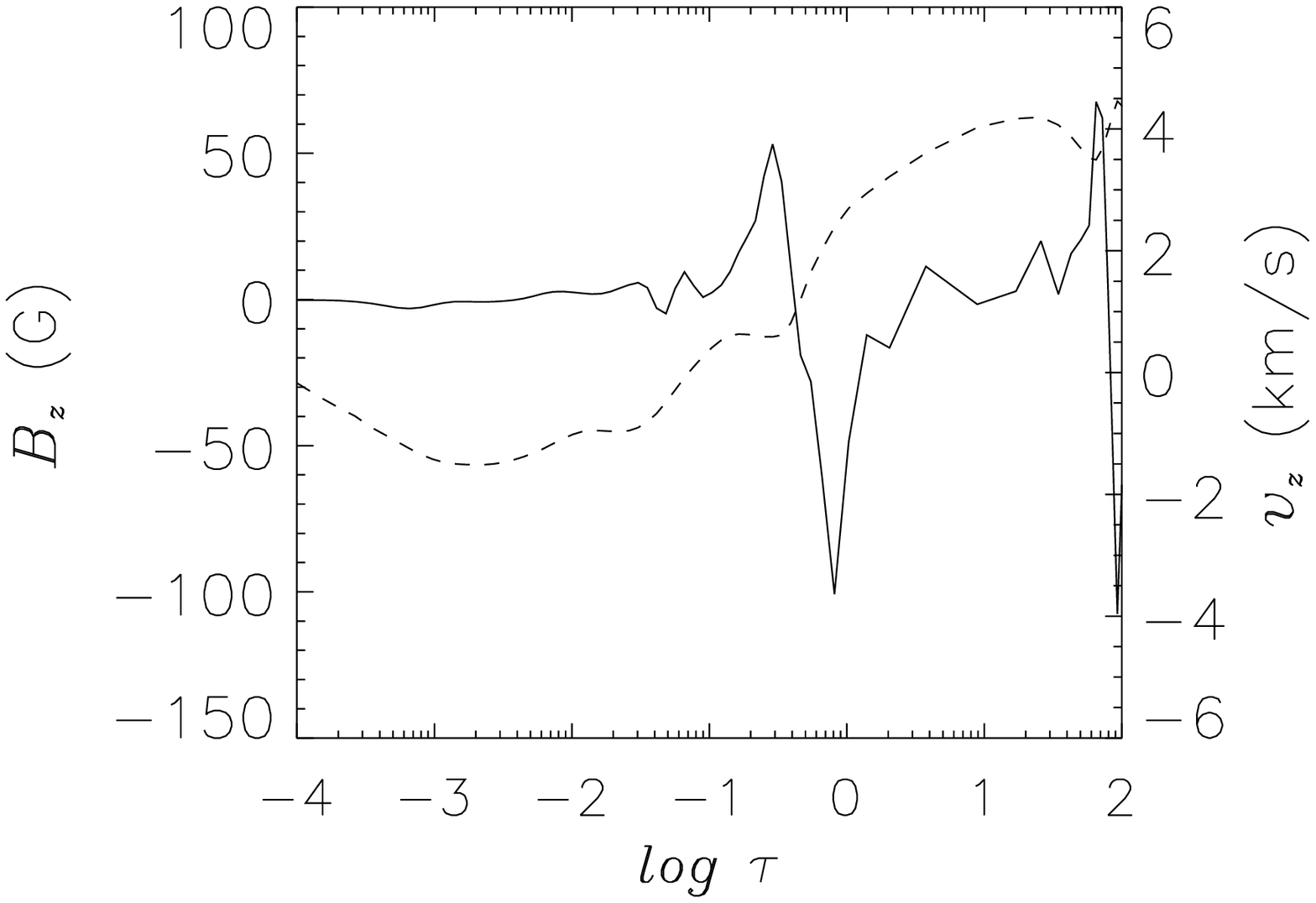}{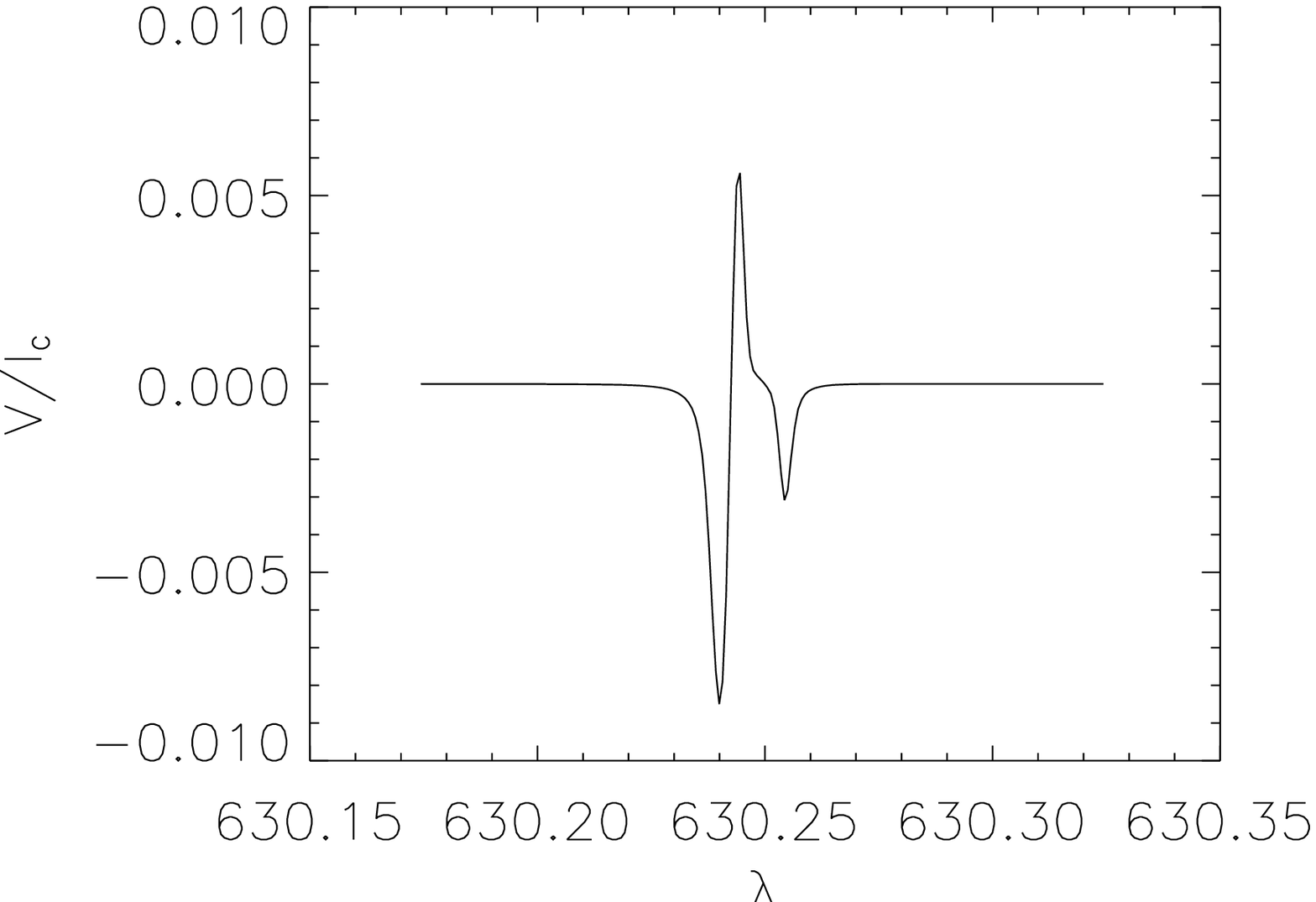}  
  \caption{{\bf (Left)} $B_z$ (solid line) and $v_z$ (dashed
    line) versus optical depth, $\tau_{500\,\mbox{nm}}$, and {\bf
      (Right)} Stokes~$V$ profile for the pixel indicated by a diamond
    in Fig. \ref{FIG:SNAPB} (\blappa$=-5.9\,$G and
    $B_{\mbox{ave}}=-1.6\cdot10^{-3}\,$G). At $\log\tau=0$ the positive and
    negative contributions to $B_{\mbox{ave}}$ have nearly cancelled
    (integrating downward).  The Stokes~$V$ signal is stronger than
    would result from a uniform $1.6\cdot10^{-3}\,$G field but is
    asymmetric.  Strong gradients lead to asymmetric profiles but also
    to $|B^{\mbox{L}}_{\mbox{app}}| \gg |B_{\mbox{ave}}|$.}
  \label{FIG:RAY}
\end{figure}

\begin{figure}
  \plotone{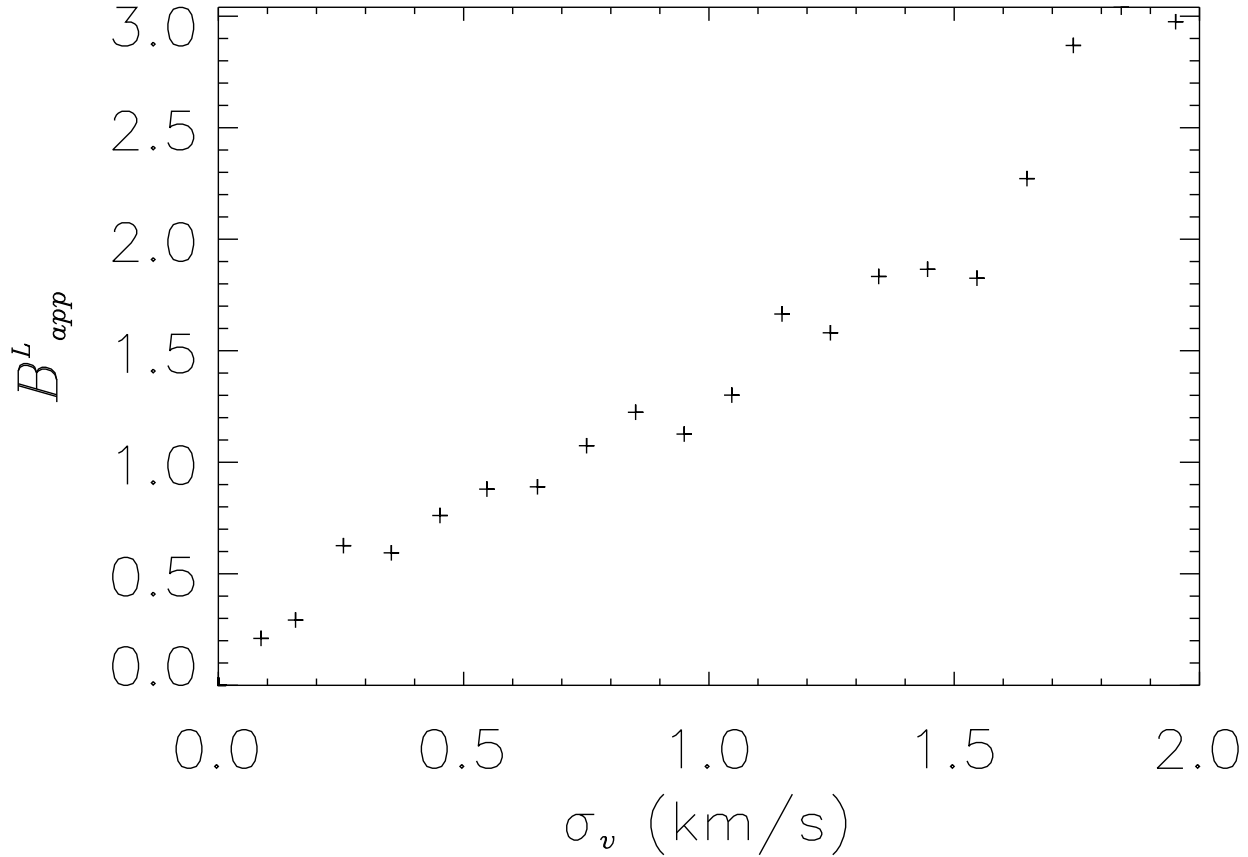}
  \caption{Average \blappa versus standard deviation of the fluctuations
of the vertical velocity along the (vertical) line-of-sight, $\sigma_v$, for all pixels with
    $|B_{\mbox{ave}}|<0.1\,$G.  Pixels are binned by $\sigma_v$ before
    averaging.  With strong velocity differences between different
    heights in the atmosphere, the total Stokes~$V$ signal increases as
    the Doppler-shifted absorption from positively and negatively oriented fields
    show less cancellation.  }
  \label{FIG:SAMI}
\end{figure}

To
understand in more detail how radiative transfer affects contribute to a peaked
\pdf, we examine a few selected $V-$profiles.  Pixels with weak $B_{\mbox{ave}}$
must be generating strong \blappa signals for \pdf(\blapp) to
become peaked.  Indeed, Fig. \ref{FIG:SNAPB} shows this is the case.
There are many pixels for which $|B_{\mbox{ave}}| < 0.1\,$G while
$|B^{\mbox{L}}_{\mbox{app}}| > 5\,$G.  On the other hand, we see that when
$|B^{\mbox{L}}_{\mbox{app}}| < 0.1\,$G, $|B_{\mbox{ave}}|$ is always less than
4~G.  In Fig. \ref{FIG:RAY}, we examine one case of how weak $B_{\mbox{ave}}$
can be associated with strong \blapp.
For $\tau \in [0.1,1]$, the vertical magnetic field takes on
values of tens of Gauss.  In this region, there are also strong
gradients (and direction reversals) for both the magnetic and velocity
fields.  In this case, because of the magnetic field reversal,
$B_{\mbox{ave}}$ is nearly zero.  However, because of the velocity gradient,
the contributions to the Stokes$-V$ profile from the positive and negative
magnetic polarities
are Doppler-shifted with respect to each other.  For this
reason, the Stokes~$V$ signal is not cancelled.  This is further
illustrated in Fig. \ref{FIG:SAMI} where we plot the mean \blappa for
all pixels with $|B_{\mbox{ave}}| < 0.1\,$G versus the strength of the
vertical velocity fluctuations.  There is a clear trend of stronger
signal with increased Doppler shifts between the different heights in
the atmosphere.  We conclude that Doppler shifts of absorption profiles are
  responsible for the peaked \pdfa from our noiseless synthetic
  \blapp.  In effect, $|V_{\mbox{tot}}|$ is some combination of
the vertical mean of $B_z$ and the vertical mean of $|B_z|$ (depending on
$v_z(z)$) and, consequently, the \pdfa of \blappa does not correspond
to \pdf$(B_z)$ from any geometrical height, optical depth, or volume. Such a failing of \blappa to accurately represent
$B_z$ cannot be captured using the Milne-Eddington approximation
(used to calibrate \blapp), which has no gradients by definition.

To examine the effect of noise on the \pdf, we consider synthetic
  Stokes~$V$ profiles with noise added at a polarization precision of \AD{$1.1\times10^{-3}$} (similar to that of the \hinodea
  observations) in determining \blapp.  The
\pdfa of this noisy synthetic observation is shown as a \AD{dotted}
line in Fig. \ref{FIG:PDFCOMPARE} \AD{and closely resembles the observational \pdfa for signals weaker than a few Gauss.  Note that,} the noise accentuates the
peak in the \pdfa even further.  In
examining Eq. (\ref{eq:vtot}) for $V_{\mbox{tot}}$ (\blappa is a nearly-linear
function of $V_{\mbox{tot}}$), we see that by taking the absolute value of the
blue and red lobes separately the effect of noise becomes the sum of two
non-negative measurement errors.
That is,
\begin{equation}
V_{\mbox{tot}}^{\mbox{measured}} = V_{\mbox{tot}}^{\mbox{true}} + |\epsilon_b| + |\epsilon_r|\,,
\end{equation}
where $\epsilon_b$ and $\epsilon_r$ are 
the measurement noise in the blue and red lobes
(e.g., $\epsilon_b\equiv\sum_{i=1}^{N_b}\epsilon_i/N$ where $\epsilon_i$
are the random variables associated with the measurement noise in each wavelength
bin).  
Assuming these two random variables, $\epsilon_b$ and $\epsilon_r$ have
Gaussian distributions, their separate \pdfsa for their
absolute values will peak at zero.  However, the \pdfa of the sum of
their absolute values will peak at a non-zero value due to
  reduced likelihood that $|\epsilon_b|$ and $|\epsilon_r|$ are small
  simultaneously: the \pdfa of the sum of two independent random
  variables is the convolution of their individual \pdfs,
\begin{equation}
P(\epsilon) =
\frac{2}{\pi\sigma_b\sigma_r}\int_0^\epsilon e^{-(\epsilon-\xi)^2/2\sigma_b^2} e^{-\xi^2/2\sigma_r^2} d\xi
\label{eq:pdf}
\end{equation}
for $\epsilon_b$,$\epsilon_r$ Gaussian and $\epsilon=|\epsilon_b|+|\epsilon_r|$.  Because of taking the absolute
values, the individual \pdfsa are zero for negative values (this sets the limits
of integration for Eq. (\ref{eq:pdf})).  Hence, their convolution is
zero at zero and peaks instead for some finite positive value.
Assuming $\epsilon_b$ and $\epsilon_r$ have identical identical
standard deviation \AD{$\sigma = \sqrt{2}\cdot2.4\,$G} (taken from \citealt
{LiKuSoNa+2008}), it can be shown that the peak in the \pdfa for
\blapp, Eq. (\ref{eq:pdf}), is given by the solution to
\begin{equation}
\frac{B}{\sigma^2}\int_0^{B/2} e^{-\xi^2} d\xi - e^{-B^2/4} = 0.
\label{eq:predict}
\end{equation} 
This predicts a peak in the \pdfa at \blapp$\approx3\,$G, %the same
%as seen in the noisy synthetic observation and
close to \AD{that} the seen in the actual observation.
  %This is what is
%seen in Fig. \ref{FIG:PDFS} when noise is added to the synthetic
%V-profiles before determining the synthetic magnetograms.  Together this
%effect and the effect of the vertical radiative transfer through a
%turbulent fluid contribute to the peak near $3\,$G seen in
%the observation.
 Consequently, adding noise leads to a
further decrease of the number of pixels with very weak field and thus
accentuates the maximum of the \pdf.  This illustrates that a monotonic \pdf$(B_z)$ is qualitatively
consistent with observations.

\begin{figure}
  \plotone{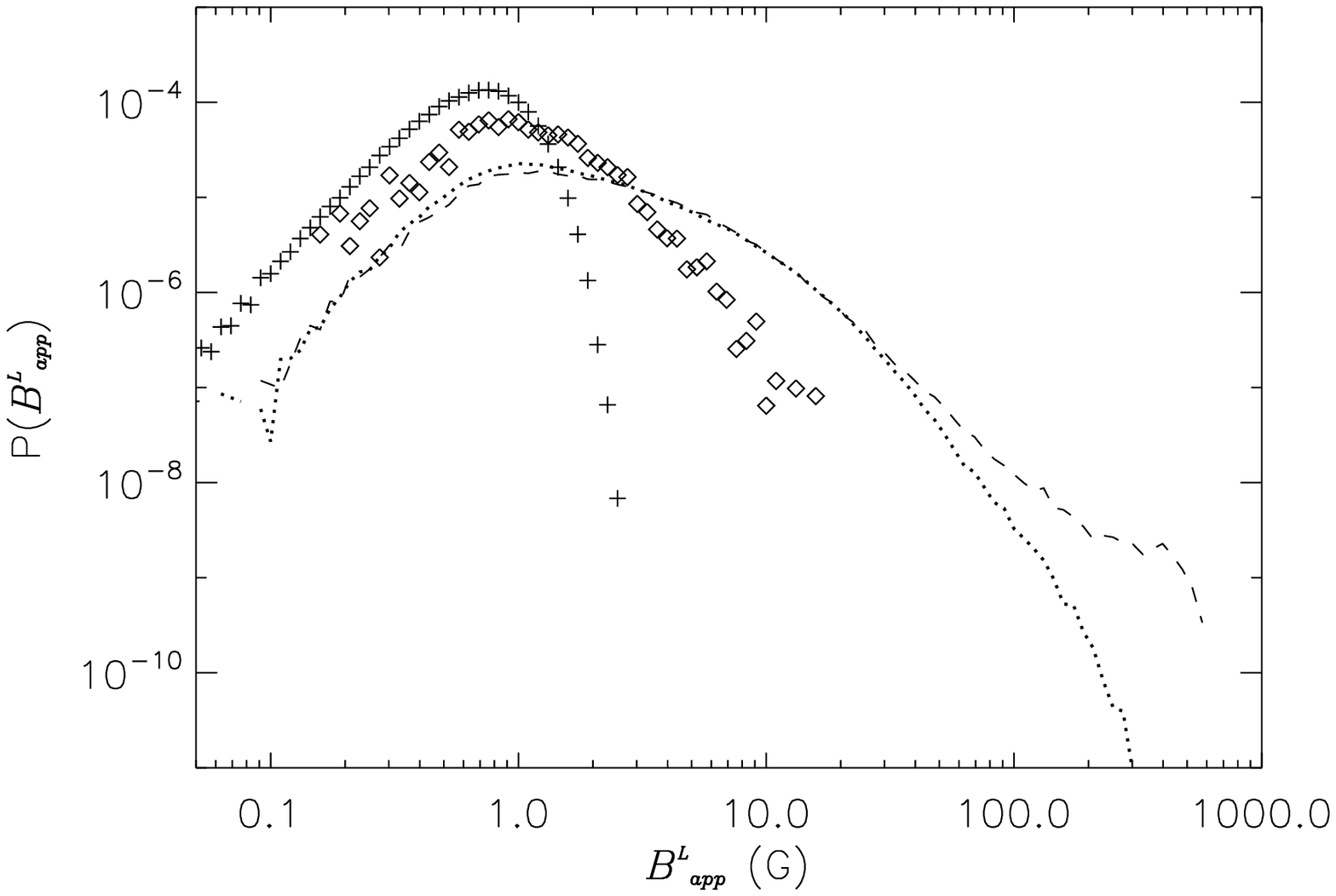} 
  \caption{\AD{\pdfsa for derived field proxies: \hinodea SP ``deep
      mode'' time series \blappa (dashed line) and \murama synthetic
      \blappa including a noise level of $3\times10^{-4}$
      (dotted). The effects of cancellation due to finite spatial
      resolution are seen in the \pdfa of the synthetic signal
      including noise as well as spatial smearing from a theoretical
      PSF and rebinning to \hinodea resolution (diamonds).  As this represents a real loss of data, the
      true mean unsigned vertical flux density cannot be calculated
      from the observational \pdf.  Also shown is the result for
      employing pure white noise with a standard deviation of $3\times10^{-4}$ for
      Stokes~$V$ (+).}}
  \label{FIG:PSFPDF}
\end{figure}

\AD{In Fig. \ref{FIG:PSFPDF}, we use the $67.2\,$s exposure ``deep
  mode'' SP time series (dashed line).  Note that for this exposure
  time, $\sigma = \sqrt{2}\cdot0.6\,$G \citep{LiKuSoNa+2008} and
  Eq. (\ref{eq:predict}) predicts that the \pdfa will peak at
  \blapp$\approx1\,$G (the actual peak is at $\approx1.2\,$G).  In
  this case, as in the ``normal mode'' case, the synthetic \murama
  \blappa with equivalent noise level (dotted line) matches the
  location of the peak and the \pdfa to the left of the peak.  In
  fact, if we generate \blappa from pure white noise for Stokes~$V$
  (standard deviation of $3\times10^{-4}$), we find its \pdfa (plus signs) predicts
  well both the location of the peak and the shape of the weak-signal
  portion of the \pdfs.  This strongly suggests that the observational
  peak is dominated by noise.}

\AD{We also find (see Fig. \ref{FIG:PSFPDF}) that cancellation of opposite polarity fields in a
  resolution element alters the \pdfa and renders it useless for
  computations of the mean unsigned flux density and other moments.
  This is evidenced by a comparison of \pdfsa from the noisy synthetic
  \murama \blappa without (dotted line) and with spatial degradation
  by a theoretical point spread function (PSF, see
  \citealt{DaGaLa+2008} for details) for \hinode's optical system
and rebinned to \hinodea pixel size
  (diamonds).  Because of the importance of the cancellation on the
  \pdfs, \pdfsa may not be used to infer the true mean unsigned
  vertical flux density.} %A more
%quantitative comparison
%\pdfsa from simulations with the observed \pdfa will be made in
%\citet{Da+2008b}.
 %They find that
 % multiplication of the
                                 %simulation's magnetic field by a
                                 %factor of 3 (crudely accounting for
                                 %the too small Reynolds numbers)
                                 %brings the observational and
                                 %simulated \pdfsa into agreement.

%The results of this section are pertinent to
%magnetogram-type observations of very weak signals to determine the
%mean unsigned vertical flux density (see, e.g,. \citealt{LiKuSoNa+2008}).  Observations
%of stronger signals analyzed with inversions for stratified
%  atmospheres may not be subject to these limitations.  In any case,
%observations of both weak and strong signals in Stokes~$V$ are limited
%by the cancellation of signal from opposite polarity fields.  This is
%addressed in the following section.

\subsection{\AD{Cancellation}}
\label{SEC:CANCEL}

Turbulence gives rise to a statistically self-similar fractal pattern
of the magnetic field (within the
inertial range) -- the field retains the same degree of
complexity of distinct structures regardless of the scale at which it is observed (see, e.g., \citealt{CoPr1992,BrPrSe+1992}).
In this section, we show that this also applies to solar surface
magnetic fields and we use this self-similarity to
estimate the portion of unsigned vertical flux unobservable at a given
resolution.  To begin with, we examine the cancellation properties of
the magnetic field itself using a series of high-resolution \murama dynamo simulations.  This illustrates how
the turbulent nature of the magnetic field limits measurement under
the sole consideration of spatial resolution and in the absence of
other observational constraints.  As it separates the statistics of
the field itself from observational constraints, the study also allows
us to extrapolate the results to realistic solar magnetic Reynolds numbers.

Extending the ideas of singularity in probability measures for
self-similar fractal fields to signed fields, \citet{ODS+92}
introduced the {\it cancellation exponent} for studying the self-similar
sign oscillations on very small scales in turbulent flows.  For our
application, their partition function, $\chi(l)$, measures the
  portion of the flux remaining after averaging over boxes of edge
  length $l$,
\begin{equation}
\chi(l) \equiv \frac{\sum_i \bigg{|}\int_{\mathcal{A}_i(l)}B_zda \bigg{|}}
{\int_{\mathcal{A}}|B_z|da} %= \frac{\Phi_{u,l}}{\Phi_u}
\label{eq:chi}
\end{equation}
where $\{\mathcal{A}_i(l)\}\subset \mathcal{A}$ is a hierarchy of disjoint subsets 
of size $l$ covering the entire domain, $\mathcal{A}$.  In our case, we call the function $\chi$ the {\it cancellation function} since it measures 
the flux cancellation at a given length-scale $l$.
If the magnetic field is self-similar (for scales much larger than the dissipation scale), we expect a power-law
\begin{equation}
\chi(l) \propto l^{-\kappa}\,,
\label{eq:ss}
\end{equation}
where $\kappa$ is called the {\em cancellation exponent}.  It
is related to the characteristic fractal dimension of the
magnetic field structures on all scales, $D_f$, by
\begin{equation}
\kappa = (d - D_f)/2
\label{eq:vsp94}
\end{equation}
where $d=2$ is the embedding Euclidean dimension of the solar surface
\citep{SVCN+02}.\footnote{Assuming the field is smooth (correlated) in
  $D_f$ dimensions and uncorrelated in the other $d-D_f$ dimensions,
  the smooth dimensions contribute to the sum of vertical fluxes
  proportional to their area while the integral of an uncorrelated
  field contributes proportional to the square root of its area
  (random process).  Eq. (\ref{eq:vsp94}) then follows
  \citep{SVCN+02}.}  An improved method to determine $\chi(l)$ using a
Monte Carlo box counting technique was proposed by
\citet{CaLaRu+1994}.  Its advantages include better counting
statistics when $l$ is a large fraction of the edge length of the
domain $\mathcal{A}$, applicability to non-square pixels, and less
sensitivity to the accidental placement of larger flux patches (e.g.,
network elements) with respect to the partitioning.  For our simulation
data, the Monte Carlo technique proved as accurate as rigid partition
boxes but led to a significant reduction of the noise in $\chi(l)$: it
averages over many partitionings and allows a more faithful
representation of the field distribution \citep{CaLaRu+1994}.  We use
this technique for the results shown below.

\begin{figure}
  \plotone{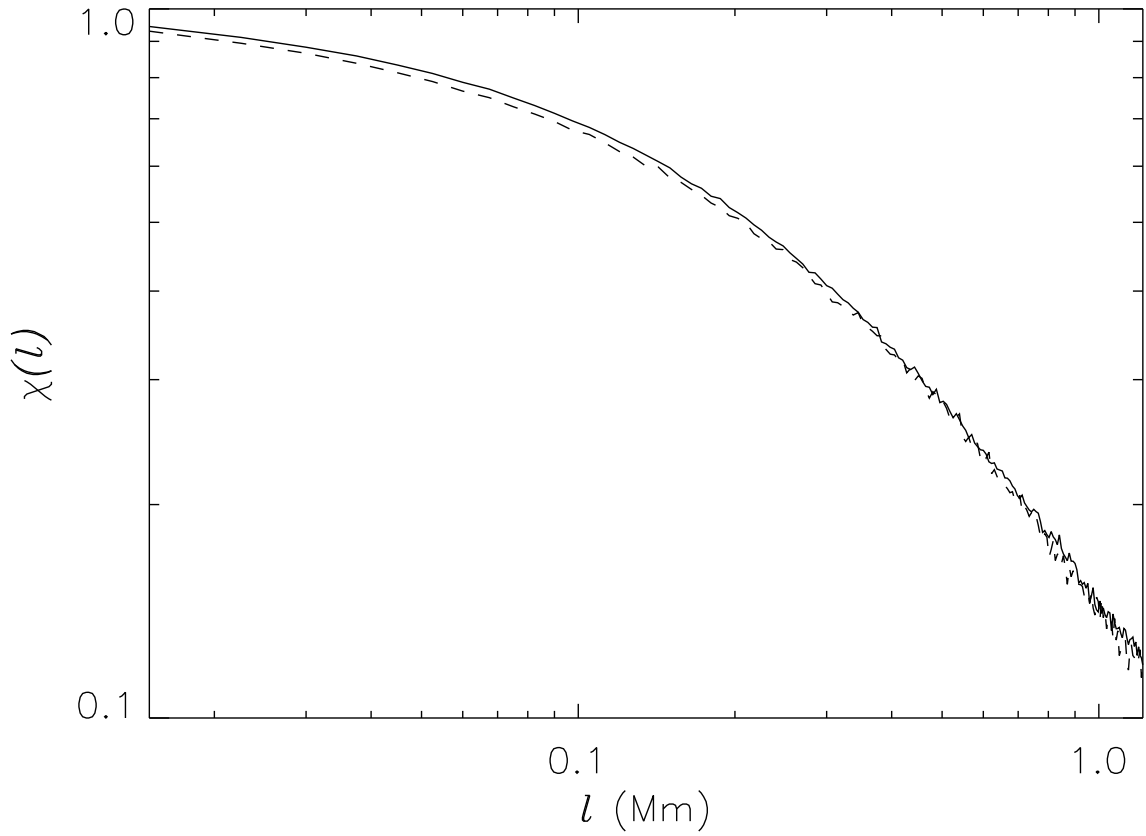}
  \caption{Cancellation function, $\chi(l)$,
    versus scale, $l$ for \runngrey: $B_z$ (solid line) and \blappa
    (simulated observation, dashed).  The two are essentially equivalent,
    suggesting that the cancellation of \blappa may be taken as a
    proxy for the cancellation of $B_z$.  %For comparison with
    %observations, \runngreya \blappa results are rebinned, degraded by
    %a theoretical \hinodea PSF and noise level
    %(asterisks). Average $\chi(l)$ from all \hinodea
    %$7\arcsec.2 \times 7\arcsec.2$ subregions with
    %mean |\blappa| below $10\,$G are shown as +'s.
 }
  \label{FIG:USE_CAEXP}
\end{figure}

The height range that corresponds (in a horizontally averaged sense) to $\log\tau\in[-2,0.1]$
(as discussed in \S\ref{SEC:PDFS}, the contribution for
  $\log\tau\in[-3.5,-2]$ to the Stokes~$V$ signal is insignificant)
is $z\in[210,300]\,$km ($z=0$ corresponds to the continuum optical
depth $\tau=1$ at $500\,$nm).  For this height range
we compute the averaged cancellation functions, $\chi(l)$, for \murama dynamo
simulations with magnetic Reynolds numbers ranging from $Re_M \approx
2000$ to $Re_M \approx 8000$. % in Fig.~\ref{FIG:EXTRAP1}.  We find no
%power-law scaling range for any of the simulations, while simple
%fractal geometry scaling from thresholding from lower $Re_M$
%{(and very different initial conditions)} simulations show such
%a range \citep{JaVoKn2003}.  One difference with this statistic to
%bear in mind, is that even for MHD simulations in simple geometries,
%$\chi(l)$ shows much less of an inertial (scaling law) range than
%other statistics like energy spectra and structure functions (see,
%e.g., \citet{PGMP05,PGHM+06}).  This could be attributed to the
%cancellation between opposite signs reducing the effective size of
%the statistical sample, for instance.
By definition, we have $\chi(l)=1$ at the resolution of the simulation
  since there are no smaller scales for the computation.  Furthermore,
 we expect dissipation
  to strongly affect $\chi(l)$ for the smallest decade of scales
  (analytically, its slope must go to zero).  Also, as our dynamo
  simulations have zero signed total flux, $\chi(4.86\,$Mm$)=0$ and we would
  expect scales down to approximately $490\,$km to be affected by this constraint.
  Only for smaller scales should we be able to observe a turbulent
  scaling.  However, very little room is left between these two
  constraints so that no clear power-law scaling is observed
  for any of the simulations (see Fig. \ref{FIG:USE_CAEXP} for one
  example).

\begin{figure}
  \plotone{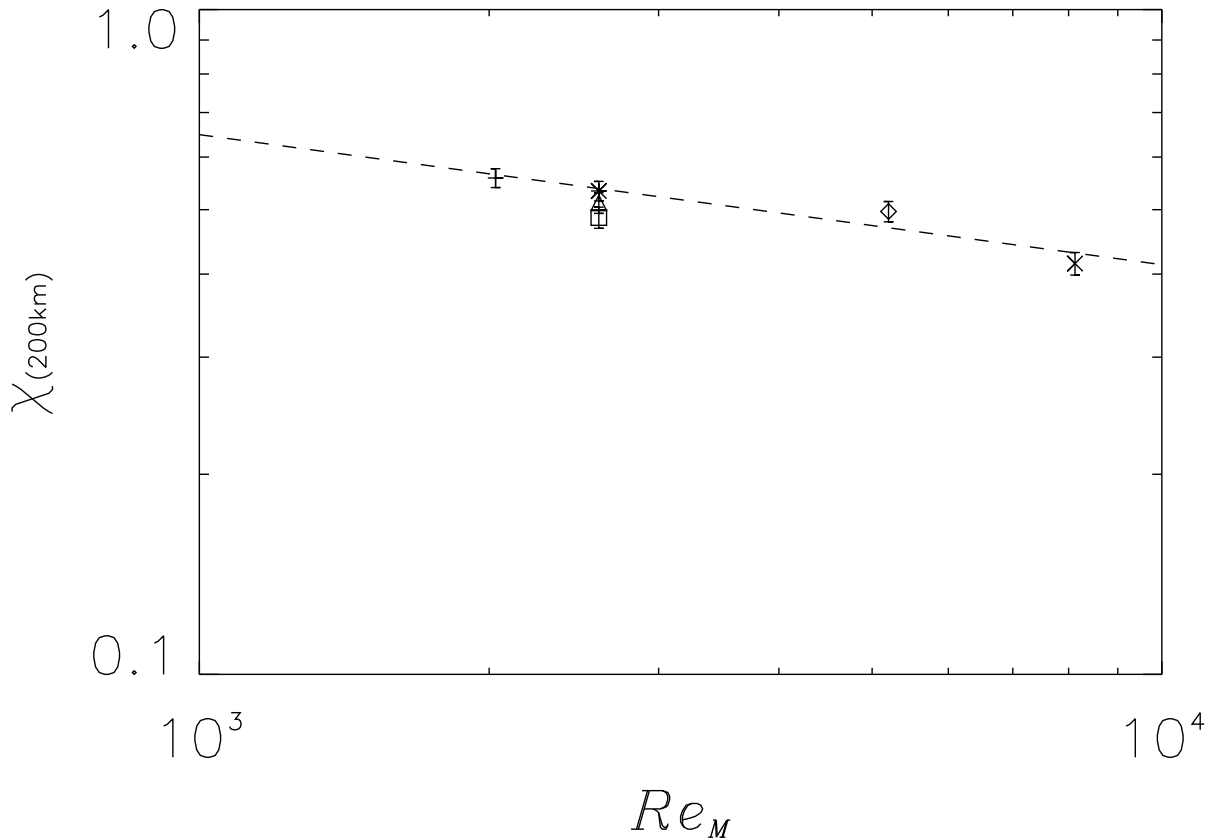}
  \caption{Portion of flux remaining at $l=200\,$km,
    $\chi(200\,$km$)$, versus magnetic Reynolds number, $Re_M$.
    Symbols are \runea (plus), \runca (asterisk), \runultraa (diamond),
    \runhypera (X), \runngreya (triangle), and \runprandtla
    (square)--see Table \ref{TABLE:RUNS}.  For fixed $l$, $\chi(l)$
    decreases with $Re_M$ and shows an
    approximate power-law relation with $Re_M$ as indicated by the
    fitted dashed line.  \runngreya and \runprandtla are not included
    in the fit, but the effect of decreased magnetic Prandtl number
    leads to reduced $\chi(200\,$km$)$.  Taking this into
    consideration, along with extrapolation to solar values,
    $Re_M\sim3\cdot10^5$, we estimate $\chi(200km)\la0.2$.}
  \label{FIG:EXTRAP2}
\end{figure}

Since the dissipation scale of magnetic energy, $l_\eta$, decreases
with increasing $Re_M$, for fixed $l$, $\chi(l)$ decreases with
increasing magnetic Reynolds number (fluctuations at smaller scales
increase the total cancellation).  This is emphasized in
Fig.~\ref{FIG:EXTRAP2}, where we plot the value of the cancellation
function for $l=200\,$km (corresponding roughly to \hinodea SP's
angular resolution of $0\arcsec.3$) versus $Re_M$.  We can fit
a power law and extrapolate to the results we would expect from a
\murama simulation at solar $Re_M$ (which must be estimated).  From
\citet{KoCr1983}, we estimate the magnetic diffusivity for
$\log\tau=0$, $\eta \sim 10^8\,$cm$^2$s$^{-1}$. The driving of the
small-scale dynamo is mainly subsurface where $\eta$ is roughly
100 times smaller ($\eta \sim 10^6\,$cm$^2$s$^{-1}$, cf.
\citealt{Sp1974}).  For an upper limit of $\chi(200\,$km$)$, we employ the more conservative
  estimate: $\eta \sim 10^8\,$cm$^2$s$^{-1}$.  Taking the forcing
  scale to be the granulation scale, $\mathcal{L} \sim 1\,$Mm, and
  using $v_{rms} \sim 3\,$km~s$^{-1}$ from the simulation, we find
\begin{equation}
Re_M \equiv \frac{\mathcal{L}v_{rms}}{\eta} \sim 3\cdot10^5.
\end{equation}
For this magnetic Reynolds number, our extrapolation yields
$\chi(200\,$km$) \sim 0.2$.  This indicates that with a perfect
observation at this spatial resolution and assuming that the
\murama simulation faithfully reproduces the solar conditions, we should multiply an
observation by a factor of 5 to obtain the true mean vertical unsigned flux
density of the quiet-Sun internetwork.  It is also suggested
by Fig.~\ref{FIG:EXTRAP2} that $\chi(200\,$km$)$ decreases with
decreasing magnetic Prandtl number, $P_M \equiv \nu/\eta$ where
$\nu$ is the kinematic viscosity and $\eta$ the magnetic diffusivity.  As the magnetic Prandtl number of the Sun
is much less than that of the simulations, we expect that
$\chi(200\,$km$) \la 0.2$.

The cancellation functions for $B_z$ and for \blappa inferred from
\runngreya are shown in Fig. \ref{FIG:USE_CAEXP}.  We see that the
  two functions are essentially equivalent.  This demonstrates an
excellent correspondence between the cancellation of the field itself
and the signal derivable from observations (excluding instrumental effects).
Therefore, we may take the cancellation of \blappa as a proxy
  for the cancellation of $B_Z$.  This we \AD{now} do.

\begin{figure}
  \plotone{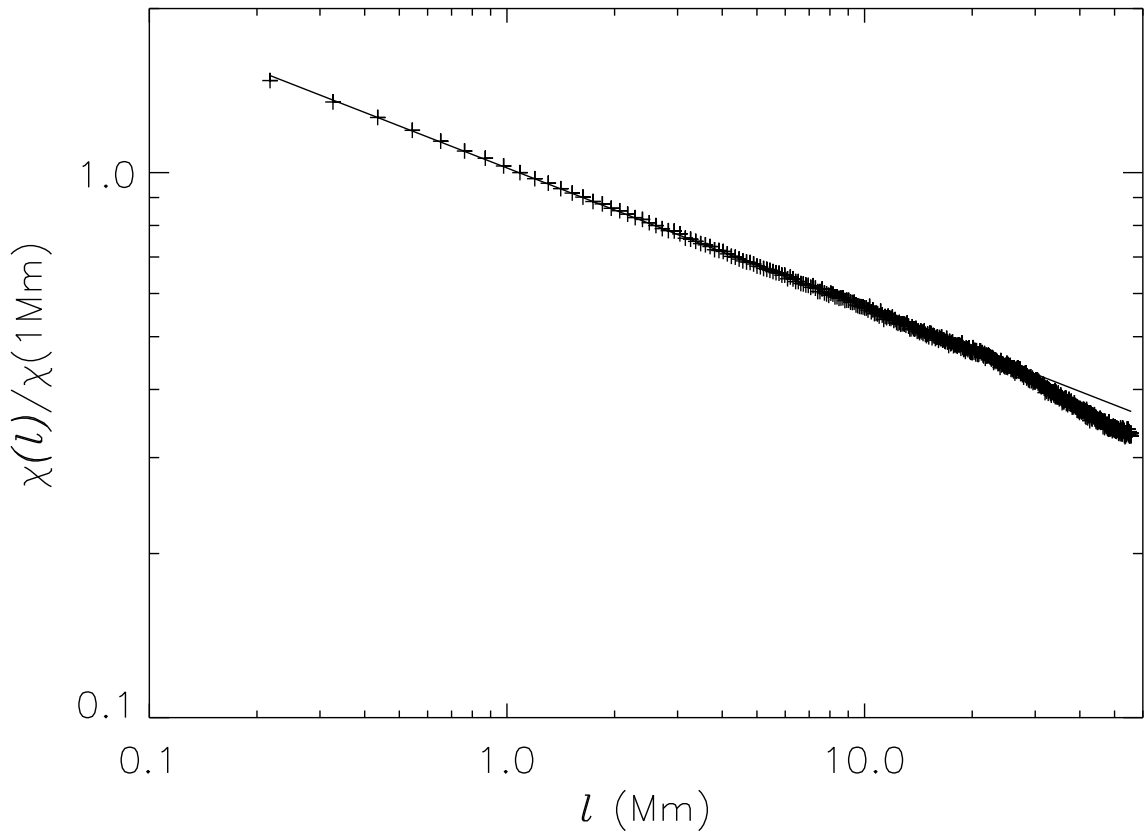}
  \caption{Normalized cancellation function, $\chi(l)/\chi(1\,$Mm$)$,
    versus scale, $l$, from \hinodea \blappa observation.  A
    self-similar power-law is abundantly clear for 2 decades of length
    scales down to the resolution limit of the observation (the fitted
    line is $k=0.26 \pm 0.01$).  This indicates both the possibility
    for self-similar extrapolation to smaller scales and that the
    smallest scales of magnetic structuring must be \AD{at least an order of magnitude smaller than $200\,$km.}}
  \label{FIG:HINODE_CAEXP}
\end{figure}

We present the normalized cancellation function,
  $\chi(l)/\chi(1\,$Mm$)$, for the \hinodea SP observation in
  Fig. \ref{FIG:HINODE_CAEXP}.  Without knowing the value of the true
  unsigned vertical flux, the denominator in Eq.~(\ref{eq:chi}), the
  value of the cancellation function can only be normalized to some
  arbitrary scale.  We find a self-similar power-law over
two decades in length scales, demonstrating the multifractal geometry of the
turbulent quiet-Sun magnetic field.  This is somewhat surprising as the dominant
granulation pattern at scales near $1\,$Mm might have been expected to
affect the cancellation scaling.  The cancellation exponent of
the scaling is $\kappa = 0.26\pm0.01$.  This exponent predicts a 20\%
increase in the observed mean unsigned vertical flux density with a
doubling of resolution in agreement with the difference in flux densities found between
ground and space-based telescopes \citep{LiKuSoNa+2008}.  Note also
that the power-law behavior holds down to the two-pixel scale.  This
is a clear indication of cancellation extending to smaller scales than
resolved by \hinodea \citep{CaBr1997,SVCV+04}.  Compare this,
  for example, to the simulation case in Fig.~\ref{FIG:USE_CAEXP}
\AD{(also see Fig.~3(b) of \citealt{SaAlEmCa2003})}
  where dissipation is seen to affect a strong turnover in $\chi(l)$
  for the smallest decade of scales.  As the observation is not so affected, we may safely
  conclude that the smallest scale of magnetic structuring is at least
  one decade smaller than the \hinodea SP resolution limit.  The scales of magnetic
  structuring in the photosphere must therefore extend to \AD{at least an order of magnitude smaller than $200\,$km.}

From Eq. (\ref{eq:vsp94}), we see that our result corresponds to
$D_f=1.48\pm0.02$ for the fractal dimension of the quiet Sun
internetwork magnetogram.  Within uncertainties, this is the
same dimension as for solar plage regions, $D_f=1.54\pm0.05$
\citep{BaScZw+1993}.  This might indicate that some similar
mechanisms are at play in solar plage and quiet Sun internetwork.  For
the cancellation exponent of
network magnetic fields, values of $\kappa\sim0.4$ \citep{LaRuCa1993}
and of $\kappa\sim0.12$ \citep{CaLaRu+1994} have been reported, but
without an estimate of the uncertainties.

Recent work has highlighted the sensitivity of fractal dimension (perimeter-area)
estimators to pixelization and resolution \citep{CrRaEr+2007}.  By
using a signed measure, however, we avoid  difficulties
inherent to fractal dimension
estimations using bi-level images in general and the perimeter-area method, specifically.
Nonetheless, we have tested the
sensitivity of the cancellation exponent to reducing our resolution by
theoretical point spread functions for apertures 1/2 and 1/4 that of the
\hinodea SOT ($50\,$cm).  We find the slope of $\chi(l)$ to be robust in these cases
for lengths exceeding 30 pixels.  There is no change in the power law for
almost one decade of length scales (3-20Mm).  We therefore conclude that
our estimation, $\kappa=0.26\pm0.01$ is robust and insensitive to
pixelization and resolution effects.  As pointed out by
\citet{LaCaRu1996}, however, because of what they call
``resolution-limited asymptotics'', different definitions of fractal
dimension can give different values at finite resolution.  For this
reason,  our value $D_f=1.48\pm0.02$ might differ from
a well-resolved perimeter-area estimate.

%{In Fig.~\ref{FIG:USE_CAEXP}, we seek to compare normalized
%  cancellation functions for our simulations and the \hinodea
%  observation.  In order to compare similar quantities, we degrade
%  synthetic profiles with the theoretical point spread function (PSF)
%  for \hinode's optical system and other effects \citep{DaGaLa+2008}.
%  For the \hinodea observation, we divide the field into
%  $7\arcsec.2 \times 7\arcsec.2$ subregions
%  (corresponding to the size of the \murama box) and consider only
%  those regions for which $\overline{|B^{\mbox{L}}_{\mbox{app}}|}<10\,$Mx~cm$^{-2}$
%  ($\overline{|B^{\mbox{L}}_{\mbox{app}}|} = 6.9\,$Mx~cm$^{-2}$ for \muram).  Both the
%  degrading from \murama resolution and the inclusion of only
%  weak-field regions bring the simulations and the observations into
%  closer agreement.  Yet, a discrepancy remains.  The simulated
%  $\chi(l)$ is a steeper function with greater curvature.  \murama
%  simulations have been shown to reproduce many of the large-scale
%  features of solar magnetoconvection, but small-scale turbulent
%  properties like the cancellation function differ from those in the
%  sun.}

Using the self-similar power law derived from
Fig. \ref{FIG:HINODE_CAEXP}, we may estimate the true mean
unsigned vertical component of the magnetic field (hereafter, ``mean unsigned vertical flux density'') in the quiet-Sun photosphere, $\Phiu$.  Below the
magnetic dissipation scale, $l_\eta$, there is no cancellation:
$\chi(l_\eta)\equiv1$.  This, together
with the self-similarity relation, Eq. (\ref{eq:ss}), gives
\begin{equation}
\Phiua = \Phiu_{l_\eta} = \Phiu_{l} \cdot
\left(\frac{l}{l_\eta}\right)^\kappa\,,
\end{equation}
where $l$ is any scale in the inertial range, $\Phiu_{l}$ is the
mean absolute value of the vertical component of the field measured at that resolution ($l$),
\begin{equation}
\Phiu_{l} \equiv \frac{\sum_i \bigg{|}\int_{\mathcal{A}_i(l)}B_zda \bigg{|}}
{\int_{\mathcal{A}}da} = \chi(l)\cdot\Phiu\,,
\end{equation}
and $\Phiua$ is given by
\begin{equation}
\Phiua \equiv \frac{\int_{\mathcal{A}}\bigg{|}B_z\bigg{|}da }
{\int_{\mathcal{A}}da}\,.
\end{equation}
\citet{LiKuSoNa+2008} report $\Phiu_{0.11}\approx11.7\,$G.
As $l\approx0.11\,$Mm (approximate \hinodea SP pixel size)
is below the SOT resolution limit, however, we rebin \blappa to
$l\approx0.22\,$Mm pixels to find $\Phiu_{0.22}\approx10.7\,$G
as the starting point of our estimate,
\begin{equation}
\Phiua \approx 10.7\,\mbox{G} \cdot
\left(\frac{0.22\,\mbox{Mm}}{l_\eta}\right)^{0.26}\,.
\label{eq:correct}
\end{equation}

Estimating the magnetic dissipation scale is not straight-forward.  As
we have shown that observationally it is unresolved, we are left to
rely on a phenomenological estimate.  Kolmogorov phenomenology
predicts (see, e.g., \citealt{F95}) $l_{\eta} \approx
\mathcal{L}Re_M^{-3/4}$ where $\mathcal{L}$ is a large characteristic
scale, such as the granulation scale.  Using $Re_M \sim
  3\cdot10^5$, derived previously, we estimate $l_{\eta} \sim 80\,$m.
  For the dissipative range, power-law scaling for $\chi(l)$ will not apply and the
  slope of the cancellation function will approach zero.  To provide a
  lower bound to the solar mean unsigned vertical field, we should
  then be conservative by ignoring any cancellation in the first
  decade of scales.  Hence, we use $l_{\eta} = 800\,$m in
  Eq. (\ref{eq:correct}) to estimate the true mean unsigned vertical
  flux density to be $\Phiua\ga46\,$G.  This means that, at a
  resolution of $200\,$km, at most one quarter of the unsigned
vertical flux is
observable.

\section{Discussion}

Our estimates suggest that three-quarters or more of the
vertical unsigned magnetic flux is cancelled at the resolution of
\hinode.  Hanle-based estimates suggest
$\left< |B| \right>\sim130\,$G \citep{TrBuShAsRa2004}\footnote{Rather than assuming that
  a turbulent magnetic field possesses a delta-function \pdfa which
  leads to the $\sim60\,$G estimate in \citet{TrBuShAsRa2004}, we take
  here the $\sim130\,$G estimate from their assumption of an
  exponential \pdf.}  while Zeeman-based estimates suggest ${\left< |B_z| \right>}\sim10\,$G
(see Table 3 in \citealt{BeGoYeChOk+2008}).  Note that even with
estimation of the cancellation, there remains almost a factor of 3 difference
between reported Hanle estimates and the Zeeman-based estimates we
present.
However, we have considered only one component of a vector quantity
while the Hanle-based estimates are sensitive to the magnitude of that
vector.  Recent observations \citep{LiKuSoNa+2008} and simulations
\citep{StReSc+2008,ScVo2008} suggest that horizontal fields are on average
a factor of 5 stronger than vertical fields.  %Even so, a mean vertical
%field strength of $\sim10\,$G combined with a mean horizontal field
%strength of $\sim50\,$G fall short of the estimates based on
%Hanle-effect depolarization from \texttt{Sr I} ($\sim130\,$G,
%\citealt{TrBuShAsRa2004}).
Therefore, our estimate of $\Phiu\ga46\,$G coupled with an even stronger mean horizontal field is
consistent with the Hanle-based estimate. Another observational discrepancy
lies in determining the mean ``location'' of the fields.
\citet{TrBuShAsRa2004} interpret scattering
polarization from molecular \texttt{C$_2$} to indicate the mean field
strength is weak ($\sim10\,$G) over the bright granules, so that the
turbulent field inferred from the Hanle measurements should be concentrated in
the intergranular lanes.  This \resp{should be compared to}
the location of strong horizontal fields not in
the lanes but near the edges of granules \citep{LiKuSoNa+2008}.  In
resolving the details over the location and strength of the mean
components of the magnetic field, future work should also address the
cancellation statistics of the horizontal field (and the
linearly-polarized Stokes signals $Q$ and $U$) as well as the effect on Stokes~$V$
presented here.

\section{Conclusion}

On the basis of surface dynamo simulations, we have demonstrated that
the \pdfa generated from the Stokes~$V$ spectra are not necessarily
equivalent in form to that of the \pdfa of the vertical component of
the underlying magnetic field.  The \pdfa for Stokes~$V$ shows a
reduction of likelihood for weak vertical
magnetic field compared to the \pdfa
of the field itself.  This effect is not due to a reduction in horizontal
resolution, but is caused by a combination of vertical radiative
transfer through a turbulent fluid (via the Doppler effect) and noise.  %If the Milne-Eddington
%approximation were a good approximation of the sun and the \pdfa for
%the vertical field strength were peaked, then a peaked \pdfa for
%Stokes~$V$ observations could also be expected.  However, as the sun
%is likely very turbulent, we expect the fields to possess strong
%gradients.
That is, any systematic sampling (by geometrical height, optical depth,
or volume) of $B_z$ from the simulation yields a monotonic \pdf, \AD{but} due to
Doppler shifts between different atmospheric heights, the Stokes $V$ signal
is not such a systematic sampling.  Consequently, the \pdfa
of Stokes $V$ field estimates do not accurately represent the \pdfa of
the actual vertical magnetic field
\AD{even in the absence of noise.  Additionally, for two
different levels of noise (``normal mode'' and ``deep mode'') we have demonstrated that the peak in the
observational \pdfa is dominated by the influence of noise.
Because of these two effects,}
a monotonic \pdfa for the field can result
in a peaked \pdfa in observations and the assumption that \pdf($B_z$)
can be uniquely derived from Stokes~$V$ observations becomes dubious.

From the cancellation function for a \hinodea observation of
  the apparent longitudinal flux density, we have demonstrated that the
  multi-fractal self-similar pattern of the quiet-Sun photospheric magnetic
  field covers two decades of length scales down to the resolution
  limit, $200\,$km.  This constitutes observational evidence that the
 the smallest scale of magnetic structuring in the photosphere is \AD{at least an order of magnitude smaller than
  $200\,$km.}  The power law also allows us to constrain the quiet-Sun true
  mean unsigned vertical flux density.  We estimate the lower bound to be
  $\approx46\,$G.  Estimates based solely on our numerical simulations
  suggest that the vertical unsigned flux at \hinode's resolution
  should be multiplied by 5 to obtain the true vertical unsigned flux
(i.e., $\sim50\,$G).
  These two results are consistent and suggest that the order of
  magnitude disparity between Hanle and Zeeman-based estimates may be
  fully resolved by a proper consideration of the cancellation
  properties of the full vector field.

{\bf Acknowledgments}

The authors would like to acknowledge fruitful discussions with
S. Solanki, A. Pietarila, and R. Cameron.  \hinodea is a Japanese
mission developed and launched by ISAS/JAXA, collaborating with NAOJ
as a domestic partner, NASA and STFC (UK) as international
partners. Scientific operation of the Hinode mission is conducted by
the Hinode science team organized at ISAS/JAXA. This team mainly
consists of scientists from institutes in the partner
countries. Support for the post-launch operation is provided by JAXA
and NAOJ (Japan), STFC (U.K.), NASA, ESA, and NSC (Norway).

{\bf Note added in proof}

\resp{We would like to point out the correlations between our
  conclusions and the works of \citet{SaAlLadeMaPi+1996} and \citet{SaAlLi2000}
  who postulate structuring of the magnetic and velocity fields on
  scales much smaller than $100\,$km. They find that synthetic
  profiles generated by 3-component Milne-Eddington atmospheres
  re-produce the observed Stokes-$V$ asymmetries found in $1\arcsec$
  resolution observations.  Though they did not estimate the
  undetected photospheric magnetic flux, the results indicated a
  significant fraction remaining undetected.  \citet{DoCeSaAlKn2006}
  assumed the quiet-Sun \pdfa can be approximated by a linear
  combination of the \pdfa inferred from Zeeman observations and a
  log-normal distribution (accounting for the observed Hanle
  depolarization).  They determined that the Hanle and Zeeman signals
  are consistent with a single \pdfa with $\langle|B|\rangle\ga100\,$G (see also
  \citealt{SaAlEmCa2003}).  \citet{SaAl2006} assuming that numerical
  simulations of magnetoconvection with no dynamo action ($20\,$km
  horizontal resolution) had achieved the asymptotic rate of magnetic
  energy dissipation, derived an estimate for the unsigned magnetic
  flux contained in unresolved scales; in our notation their finding
  is $\chi(100\,$km$)\sim0.5$ while our extrapolation estimates
  $\sim0.36$. Finally, \citet{SaAl2008} using observational data from
  various sources plot $\langle|B_z|\rangle_l$ versus $l$ (their Fig. 1).  The
  data are compared to a line, the slope of which corresponds to
  $\kappa=1$, i.e., the result for white noise (\citealt{VSP+94}, also
  set $D_f=0$ in Eq. (\ref{eq:vsp94})).  There is, however, a large
  scatter about this line suggestive of either a large uncertainty in
  $\kappa$ or an element of randomness in the calibration issues
  between the various data used.}

\clearpage

%% the file, uncomment the \plotone, \plottwo, and
%% If you need a layout that cannot be achieved with \plotone or
%% \plottwo, you can invoke the  package directly with the
%% \includegraphics command or use \plotfiddle. For more information,
%% please see the tutorial on "Using Electronic Art with " in the
%% documentation section at the  Web site,
%% http://www.journals.uchicago.edu/AAS/AASTeX.
%%
%% The examples below also include sample markup for submission of
%% supplemental electronic materials. As always, be sure to check
%% the instructions to authors for the journal you are submitting to
%% for specific submissions guidelines as they vary from

%% Here we use \plottwo to present two versions of the same figure,
%% one in black and white for print the other in  color
%% for online presentation. Note that the caption indicates
%% that a color version of the figure will be available online.
%%

%% This example uses \plotone to include an  file scaled to
%% 80% of its natural size with \epsscale. Its caption
%% has been written to indicate that additional figure parts will be
%% available in the electronic journal.

%\begin{figure}
%  \plotone{extra_caexp}
%  \caption{Cancellation function, $\chi(l)$ versus length scale, $l$,
%    for \murama $B_z$ averaged over $z \in [210,300]\,$km.  Dashed
%    lines indicate the tangents at $l=0.5\,$Mm and illustrate that the
%    curves do not posses a significant power-law scaling range.  Note
%    also the strong curvature for the smallest decade of scales.
%    Colors are \runea (red), \runca (black), \runultraa (blue),
%    \runhypera (green), \runngreya (grey), and \runprandtla
%    (cyan)--see Table \ref{TABLE:RUNS}.  For fixed $l$, $\chi(l)$
%    decreases with $Re_M$.  }
%  \label{FIG:EXTRAP1}
%\end{figure}

%\clearpage

\begin{deluxetable}{lccc}
\tablewidth{0pt} \tablecaption{Summary of \murama simulation runs:
  shown are grid points, horizontal resolution, and magnetic Reynolds
  number, $Re_M$. All runs except \runngreya utilize grey
    radiative transfer.  In \runngrey, opacity binning with 4 bins
\citep{VBS04} has been used to provide non-grey radiative transfer.  For all simulations no physical
   viscosity is imposed.  Rather, numerical dissipative effects lead
  to an effective kinetic Reynolds number, $Re$ \citep{VSS+05}.  To obtain a
    lower value of $P_M = Re_M/Re$, \runprandtla uses the magnetic
    diffusivity used in \runca but at a higher resolution, hence
   higher $Re$.}
\tablehead{ \colhead{Simulation} &
  \colhead{Computational Grid} & \colhead{Horizontal Resolution} &
  \colhead{$Re_M$}}
\startdata
\runea & $540\times540\times140$ & $9\,$km & $\approx 2000$ \\ %\hline
\runca & $648\times648\times140$ & $7.5\,$km & $\approx 2600$ \\ %\hline
\runngreya & $648\times648\times140$ & $7.5\,$km & $\approx 2600$ \\ %\hline
\runprandtla & $972\times972\times200$ & $5\,$km & $\approx 2600$ \\ %\hline
\runultraa & $972\times972\times200$ & $5\,$km & $\approx 5200$ \\ %\hline
\runhypera & $1215\times1215\times350$ & $4\,$km & $\approx 8100$ \\ %\hline
\enddata
%%%\tablenotetext{a}{Star}
%%% You can append references to a table using the \tablerefs command.
\label{TABLE:RUNS}
\end{deluxetable}

\end{document}